\documentstyle[aps,prb,multicol,psfig]{revtex}
\begin{document}
\draft
%\preprint{mpi-pks/9612016}
\title
{The ground state of the Kondo model with large spin}

\author{Shun-Qing Shen}

\address
{Max Planck Institute for Physics of Complex Systems,
Bayreuther Strasse 40, Haus 16, D- 01187 Dresden, Germany}

\date{Received: December 26, 1996}
\maketitle

\begin{abstract}
In this paper, we prove that the ground state of the Kondo model with large
spin is nondegenerate, apart from a SU(2) spin degeneracy
in the  case of half filling. The ground state spin is found for the system, and  
the energy level orderings are discussed. Finally, the existence of
ferrimagnetism in some cases is proved. 
\end{abstract}
\pacs{PACS numbers: 75.10.Lp 75.30.Mb, 75.50.Gg}

%\narrowtext
\begin{multicols}{2}
%\newpage
%\narrowtext

The Kondo models, or single and lattice impurity models, are one of the most
challenging subjects in strongly correlated systems 
\cite{lee,coleman,ludwig}. 
Recent developments of reflection positivity technique in the spin space make
it possible to establish some rigorous results for the half-filled strongly
correlated electron systems
\cite{Lieb89,Ueda92,Shen94,Yanagisawa95,Shen96a,Shen96b}.
Theory of reflection positivity in the spin space for the single- and
multi-channel Kondo models with spin 1/2 was developed recently and a series
of rigorous results on the ground state properties were proved
\cite{Shen96a,Shen96b}. However some materials 
are described by the Kondo models with large
spin, such as $(La_{1-x}X_x)MnO_3$ with $X= Ba, Ca, Sr$ etc. and the localised
spin $s= 3/2$ \cite{Inoue95}. Due to more degrees of freedom in the case
of large spin than in the case of spin 1/2, usually 
it is very hard to extract
rigorous results for those systems. As a generalisation of the  theory
for the Kondo model with spin 1/2, we will investigate the Kondo model with large
spin and provide some rigorous results on the ground state of the Kondo model
in the case of half filling.

Let us first write down the Hamiltonian we will investigate:
\begin{eqnarray}
H & = & \sum_{\langle ij \rangle \in \wedge,\sigma} t_{ij} c^{\dagger}_{i,\sigma}c_{j,\sigma}
 + \sum_{i\in \wedge} U_i (n_{i, \uparrow} - {1\over 2}) (n_{i, \downarrow}
-{1\over 2}) \nonumber \\
 & + & \sum_{i\in \wedge} J_i {\bf S}_i \cdot {\bf S}_{ci}+ 
\sum_{ij \in \wedge_d} K_{ij} {\bf S}_i \cdot {\bf S}_j,
\end{eqnarray}
where $c^{\dagger}_{i,\sigma}$ and $c_{i,\sigma}$ are the creation
and annihilation operators for the conduction electron (c-electron) 
at site $i$ with spin
$\sigma (= \uparrow,\downarrow)$ and $n_{i,\sigma} =
c^{\dagger}_{i,\sigma}c_{i,\sigma}$. ${\bf S}_i$ is the localised 
spin operator with spin
$s_i$ at site $i$. ${\bf S}_{ci} = \sum_{\sigma,\sigma'}c^{\dagger}_{i, \sigma}
\bbox{\sigma}_{\sigma,\sigma'} c_{i, \sigma'}/2$, and $\bbox{\sigma}_{\alpha}$
($\alpha = x, y, z$) are
the Pauli matrices. The model is defined on a bipartite lattice
$\wedge$ with the site numbers $N_A$ and $N_B$ of the two sublattices $A$ and
$B$. $t_{ij} = t_{ji}$ is possibly non-zero only when $i$ and $j$ belong to
two different sublattices. The lattice $\wedge$ is connected by the hopping
terms $\{ t_{ij}\}$, which implies that for any two sites $k$ and $l$ on
$\wedge$, we can always find a sequence $\{ (k, i_1), (i_1, i_2), \cdots,
(i_n, l)\}$ such that $t_{ki_1}t_{i_1i_2}\cdots t_{i_nl} \neq 0$. $\wedge_d$ is
the distribution of localised spins on $\wedge$. 
If we regard the same site for conduction electron and localised spin as two
independent 
sites in a generalised lattice, the site $i$ for conduction electron
and the site $i$ for localised spin belong to the same sublattice when $J_i
<0$, and to two different sublattice when $J_i >0$. In the case the
generalised lattice can be still regarded as a generalised bipartite lattice.
Assume $K_{ij}J_iJ_j \leq 0$ if  $i$ and $j$
belong to the same sublattice, and $K_{ij}J_iJ_j \geq 0$ if  $i$ and $j$
belong to two different sublattices. When $J_i$ and $J_j$ are constant or
have the same sign, $K_{ij} \leq 0$ if  $i$ and $j$
belong to the same sublattice and $K_{ij} \geq 0$ if  $i$ and $j$
belong to two different sublattices as in the antiferromagnetic Heisenberg
model on a bipartite lattice. 
This condition guarantees that the last term in Eq.(1) does not introduce any
frustration for the generalised bipartite lattice. Physically, it is
completely absent of frustration in this case.
The model is reduced to the Hubbard model when $\wedge_d=0$, i.e., there is no
magnetic impurity or localised spin in the systems.
The structure function of the bipartite lattice $\wedge$ is defined
as  $\epsilon(i) = 1$ if $i \in A$ and $-1$ if $i\in B$. 

The main results are summarised as follows:

\noindent
{\bf Theorem}: Assume the model in Eq.(1)  with $t_{ij}$ and $K_{ij}$ is defined on a
connected bipartite lattice $\wedge$ with sublattice sites $N_A$ and $N_B$ and with the
distribution of localised spins $\wedge_d$. 
All $U_i>0$ and $J_i\neq 0$ and the number of conduction electrons is $N_e =
N_{\wedge} = N_A + N_B$. The signs of $K_{ij}$ and $J_i$ satisfy the previously stated condition. Denote the lowest energy state in the subspace
decomposed by the z-component of the total spin $S_{tot}^z$ by $\vert\Psi\rangle$.\\
i). The state $\vert\Psi\rangle$ is non-degenerate in each  subspace of 
$S_{tot}^z$. The ground state of Eq. (1)  is
unique apart from a ($2S^0_{tot} + 1$)-fold spin degeneracy. \\
ii). The total spin
$S_{tot}$ in the  lowest energy state is  
\begin{equation}
S_{tot} = \left\{
\begin{array}{ll}
S^0_{tot},&\makebox{if $\vert S_{tot}^z \vert \leq S^0_{tot}$}; \\
\vert S_{tot}^z \vert, & \makebox{otherwise}
\end{array}
\right. 
\label{b}
\end{equation}  
where 
\begin{equation}
S^0_{tot} = \left \vert {1\over 2}\sum_{i\in\wedge}\epsilon(i)
- \sum_{i \in \wedge_d}s_i {J_i \over \vert J_i \vert}\epsilon(i)\right \vert. 
\end{equation}
iii). The spin-spin correlation functions obey,
\begin{eqnarray}
\langle\Psi\vert{\bf S}^+_{ci}\cdot{\bf S}^-_{cj}\vert\Psi\rangle
&=&\epsilon(i)\epsilon(j)C_{ij};\nonumber \\
\langle\Psi\vert{\bf S}^+_{i}\cdot{\bf S}^-_{j}\vert\Psi\rangle
&=&\epsilon(i)\epsilon(j)F_{ij};\\
\langle\Psi\vert{\bf S}^+_{ci}\cdot{\bf S}^-_{j}\vert\Psi\rangle
&=&-\epsilon(i)\epsilon(j)\frac{J_{\perp}}{\vert J_{\perp}\vert}G_{ij}, 
\nonumber
\end{eqnarray} 
where $C_{ij}$, $F_{ij}$ and $G_{ij}\geq 0$ if $U_i\geq 0$, 
and $>0$ if all $U_i>0$.\\
iv). When all $U_i=0$, (at least one of) the ground state(s) (if degenerate) 
has the total spin as that in Eq. (\ref{b}).

Before we present the proof, several remarks or corollaries are made:

1). In the case of all $J_i>0$ or $<0$ and $s_i =s$, suppose that 
$N_{Ad}$ spins are on
the sublattice $A$, and $N_{Bd}$ spins on the sublattice $B$, the total spin
is
\begin{equation}
S^0_{tot} =\left \vert {1\over 2}(N_{A} - N_{B}) - {J\over \vert J \vert} s
(N_{Ad} -N_{Bd})\right \vert.
\end{equation}
In the Kondo lattice case, $N_A = N_{Ad}$ and $N_{B} =
N_{Bd}$, and the total spin is
\begin{equation}
S_{tot}^0 = \left (s - {1\over 2} {J\over \vert J\vert}\right )
\vert N_A -N_B\vert.
\end{equation}
When all $s_i = 1/2$, we recover the result for the case of spin 1/2 
\cite{Shen96a,Shen96b}. 
The ground state is a 
singlet only when $N_A =  N_B$, or $s=1/2$ and $J>0$. When
$N_A \neq N_B$ and $(N_A -N_B)/(N_A + N_B) \neq 0$ when the system becomes
sufficiently large, we obtain a state
with ferromagnetic long-range order.

2). Theorem (iii) indicates that strong antiferromagnetic correlations
exist between the conduction electrons or localised spins. A direct
corollary is that the antiferromagnetic correlation is always stronger than
the ferromagnetic correlation. For example on a cubic lattice,
\begin{equation}
\langle {\bf S}^+_Q \cdot{\bf S}^-_Q\rangle =\max\{
\langle{\bf S}^+_q\cdot{\bf S}^-_{-q} \rangle\},
\end{equation}
where ${\bf S}^+ = {1\over \sqrt{N_{\wedge}}}\sum_{i\in \wedge}{\bf S}^+_{i}
e^{i\vec{q}\cdot \vec{r_i}}$ and $Q = (\pi,\pi, \cdots)$. If the state
possesses ferromagnetic long-range order, it must also possess
antiferromagnetic long-range order. In other words, the ferromagnetism in
the case of $N_A \neq N_B$ is, strictly speaking, the ferrimagnetism.

3). The energy  level orderings can be obtained from Theorems (i-ii) and the
SU(2) symmetry with a variational principle: 
denote $E(S_{tot}^z)$ the lowest energy state with $S_{tot}^z$.
As $\vert\Psi(S^z_{tot})\rangle$
is the lowest energy state with $S_{tot}^z$, we can construct an eigenstate
$(\sum_{i\in \wedge}{\bf S}_{ci}^- + \sum_{i\in\wedge_d}{\bf
S}_i^-)\vert\Psi(S_{tot}^z)\rangle$ (suppose $S_{tot}^z
\geq 1$) due to the spin SU(2) symmetry.  This state has its z-component of
the total spin $S_{tot}^z-1$ and the
eigenvalue $E(S_{tot}^z)$.  Meanwhile it has the same total spin as in 
$\vert\Psi(S^z_{tot})\rangle$. In the variational principle the lowest energy
state in the subspace of $S_{tot}^z-1$ should not be higher than 
$E(S_{tot}^z)$. If the lowest energy in the subspace $S_{tot}^z-1$ is equal to
the lowest energy in the subspace $S_{tot}^z$, the lowest energy state or one
of the states if degenerate with $S_{tot}^z-1$
must has the same total spin in the state  with $S_{tot}^z$. Since the lowest
energy states in each subspace are non-degenerate when all $U_i>0$ and we
assume $S_{tot}^z \geq 0$, we have
\begin{eqnarray}
E(S_{tot}^z) &=& E(-S_{tot}^z);\\
E(S_{tot}^z) &<& E(S_{tot}^z + 1) ~ \makebox{ if $S^z_{tot} \geq S_{tot}^0$
};\\
E(S_{tot}^z) &=& E(S_{tot}^0) ~ \makebox{ if $0\leq S^z_{tot} \leq S_{tot}^0$
}.
\end{eqnarray}
For $S^z_{tot}\leq 0$, the energy level ordering can be obtained from
Eqs. (8-10). 
This is similar to that of the Heisenberg model \cite{Lieb62}. It is worth of
mentioning that in the case of $U_i\geq 0$ we have to  use ``lower or equal''
instead of ``lower than'' if we could not determine whether the lowest energy
state in each subspace is non-degenerate or not

4). In the one-dimensional chain of the Kondo lattice with all $J_i>0$ or
$<0$,  Theorems (ii) and (iii) on the total spin and spin-spin correlation
functions are the same as those in a two-chain spin-ladder system. This
coincides with 
the analysis on the resemblance of these two systems by White and
Affleck \cite{White96}.

5). When the system has no localised spin or magnetic impurity,
$N_{Ad}=N_{Bd}=0$ and the model in Eq.(1) is reduced to a Hubbard
model. Theorem (i-iv) hold for the Hubbard model. In the case, one can regard
the spin formula in Eq. (2)
as a generalisation of Lieb's theorem for the Hubbard model \cite{Lieb89}. The
conditions of even number of the lattice sites and $S_{tot}^z=0$ in Lieb's
theorem are
removed. As a corollary, the energy level ordering for the Hubbard model in
the case of half filling is the same as shown in
Eqs. (8-10).

The main steps to prove this theorem are 
1). to express the spin operator in a multi-fermion representation;
2). to define the generalised bipartite lattice after we consider the
localised spins; 3) to introduce a complete and orthonormal
set of basis for the system; 4) to prove the positive definiteness or
semidefiniteness of the ground state on the chosen basis \cite{Note1}; and 5) to prove
(iii) and (iv) by utilising 
the positive definiteness of the ground state and the known theorems on a
positive definite state.

Except for the reflection positivity approach in the spin space as in the case
of spin $1/2$ \cite{Shen96a,Shen96b}, 
the key trick used in this paper is to express the localised spin
operator ${\bf S}_i$ as the summation of $2s_i$ spin $1/2$ operators in a
multi-fermion representation (d-fermion) and these $2s_i$ spins are coupled
ferromagnetically. Completely ferromagnetic coupling could be realized by introducing a strong ferromagnetic 
coupling limit between these spins in the Hamiltonian. The spin operator is
written as
\begin{equation}
{\bf S}_i = \sum_{\alpha_i = 1}^{2s_i} d^{\dagger}_{i\alpha_i, \sigma}
{\bbox{\sigma}_{\sigma, \sigma'} \over 2} d_{i\alpha_i,\sigma'},
\end{equation}
with the restriction that each site  $i\alpha_i$ is singly occupied by
$d$-fermions. The effective Hamiltonian in terms of d-fermions is
\begin{eqnarray}
&H' & =  \sum_{\langle ij \rangle \in \wedge} t_{ij} c^{\dagger}_{i,\sigma}c_{j,\sigma}
 + \sum_{i\in \wedge} U (n_{i, \uparrow} - {1\over 2}) (n_{i, \downarrow}
-{1\over 2}) \nonumber \\
 & + & J \sum_{i, \alpha_i} (d^{\dagger}_{i\alpha_i, \sigma}
{\bbox{\sigma}_{\sigma, \sigma'} \over 2} d_{i\alpha_i,\sigma'})  \cdot 
(c^{\dagger}_{i, \sigma}
{\bbox{\sigma}_{\sigma,\sigma'} \over 2} c_{i, \sigma'})\nonumber \\
&+& \sum_{ij \in \wedge, \alpha_i,\alpha_j} K_{ij}  
(d^{\dagger}_{i\alpha_i, \sigma}
{\bbox{\sigma}_{\sigma, \sigma'} \over 2} d_{i\alpha_i,\sigma'}) \cdot
 (d^{\dagger}_{j\alpha_j, \sigma}
{\bbox{\sigma}_{\sigma, \sigma'} \over 2} d_{j\alpha_j,\sigma'})\nonumber \\
&-& \sum_{i\in \wedge} \lambda_i[(
 \sum_{\alpha_i} d^{\dagger}_{i\alpha, \sigma}
{\bbox{\sigma}_{\sigma, \sigma'} \over 2} d_{i\alpha,\sigma'})^2 - s_i(s_i + 1)],
\label{c}
\end{eqnarray}
where all $\lambda_i$ are positive \cite{Note2}. We shall take 
all $\lambda_i \rightarrow \infty$ at the last step (as a matter of fact
our theorem is true for any finite $\lambda_i$), and this guarantees that any
deviation from 
\begin{equation}
\langle \Psi\vert\left ( \sum_{\alpha_i =1}^{2 s_i}d^{\dagger}_{i\alpha_i, \sigma}
{\bbox{\sigma}_{\sigma, \sigma'} \over 2} d_{i\alpha_i,\sigma}\right)^2 \vert
\Psi\rangle = s_i(s_i +1)
\end{equation}
will lead to the divergence of the eigenvalues of energy in Eq. (12). 
Eq. (13) is an alternative
expression of 
\begin{equation}
{\bf S}_i^2\vert\Psi\rangle = s_i (s_i +1)\vert \Psi\rangle
\end{equation}
in the multi-fermion representation. 
In the limit of $\lambda_i \rightarrow +\infty$, $H'$ in Eq. (12) is equivalent to $H$
in Eq. (1). 

The definition of the generalised bipartite lattice is very crucial for
utilising the 
reflection positivity technique in the spin space. The generalised
bipartite lattice for $H'$ is defined as follows: 
1). The sublattice $A$ ($B$) for c-electrons belongs to  the
generalised sublattice  $\cal{A}$ ($\cal{B}$);
2). The site $i\alpha_i$ for $d$-fermions belongs to the generalised sublattice
$\cal{A}$ ($\cal{B}$) if $i$ belongs to the sublattice $A$ ($B$) when $J_i <0$ and to the
generalised sublattice   $\cal{B}$ ($\cal{A}$) when $J_i > 0$. At the same site $i$
all $i\alpha_i$ with different $\alpha_i$ belong to the same generalised
sublattice. Additionally,  the sites of the generalised lattice which are
connected by  $K_{ij}$ belongs to the same sublattice when $K_{ij}<0$ and to
two different sublattices when $K_{ij}>0$ according to the condition on the
signs of $J_i$ and $K_{ij}$.

According to this generalised bipartite lattice, the transformation operator {\bf
T} for the partial particle-hole transformation is introduced as \cite{Pu94}
\begin{equation}
{\bf T} = \prod_{i\in \wedge}( c_{i\uparrow} -
\epsilon(i)c^{\dagger}_{i\uparrow})\prod_{i\in\wedge_d}\prod_{\alpha_i = 1}^{2
s_i}\left (d_{i\alpha_i\uparrow} + {J_i\over \vert
J_i\vert}\epsilon(i)d^{\dagger}_{i\alpha_i\uparrow}\right ).
\end{equation}
Under this transformation, we have
\begin{eqnarray}
{\bf T} c_{i\uparrow} {\bf T}^{\dagger} &=& (-1)^{N_t}\epsilon(i)c^{\dagger}_{i\uparrow};\\
{\bf T} c_{i\downarrow} {\bf T}^{\dagger} &=& (-1)^{N_t}c_{i\downarrow};\\
{\bf T} d_{i\alpha_i\uparrow} {\bf T}^{\dagger} &=& 
(-1)^{N_t +1}{J_i\over\vert J_i \vert}
\epsilon(i)d^{\dagger}_{i\alpha_i\uparrow};\\
{\bf T} d_{i\alpha_i\downarrow} {\bf T}^{\dagger} &=& (-1)^{N_t}
d_{i\alpha_i\downarrow},
\end{eqnarray}
where $N_t$ is the total number of the generalised bipartite lattice sites.

The set of basis for $H'$ in the case of half filling we choose is $\{ {\bf T} \vert
\phi^{\uparrow}_{\alpha} \rangle \otimes \vert \phi^{\downarrow}_{\beta}
\rangle \}$. $\{\vert \phi^{\sigma}_{\alpha} \rangle \}$ is a real, complete and
orthonormal set of basis for $N_0$ c-electrons and d-fermions with spin $\sigma$, which is
expressed as $\phi^{\sigma}_{\alpha} =
\prod_{i\in \alpha} c^{\dagger}_{i\sigma}\prod_{i\in\alpha_i}d_{i\alpha_i \in
\alpha}^{\dagger} \vert 0\rangle$ (the order of $c$ and $d$ operators will
cause an undetermined phase $-1$, which does not affect our final result). The
single occupancy of $d$-fermion at site $i\alpha_i$ is
realized by choosing all  $\{\alpha,\beta\}$ such  that
\begin{equation}
n_{i\alpha_i\uparrow}\vert
\phi^{\uparrow}_{\alpha} \rangle \otimes \vert \phi^{\downarrow}_{\beta}
\rangle =  n_{i\alpha_i\downarrow} \vert
\phi^{\uparrow}_{\alpha} \rangle \otimes \vert \phi^{\downarrow}_{\beta}
\rangle
\end{equation}
for all $i$ and $i\alpha_i$. Each   $\{ {\bf T} \vert
\phi^{\uparrow}_{\alpha} \rangle \otimes \vert \phi^{\downarrow}_{\beta}
\rangle \}$ contains $N_0$ c-electrons and d-fermions with spin down and $N_t
- N_0$ c-electrons and d-fermions with spin up. The total number of c-electrons and d-fermions in  $\{ {\bf T} \vert
\phi^{\uparrow}_{\alpha} \rangle \otimes \vert \phi^{\downarrow}_{\beta}
\rangle \}$ is $N_t$. Consider now the constraint of single occupancy of
d-fermions. The number of c-electrons in this basis is $N_e = N_{\wedge}$. The
z-component of the total spin is
\begin{equation}
{\bf S}_{tot}^{z} {\bf T} \vert
\phi^{\uparrow}_{\alpha} \rangle \otimes \vert \phi^{\downarrow}_{\beta}
\rangle  = {1\over 2}(N_t -2N_0)   {\bf T} \vert
\phi^{\uparrow}_{\alpha} \rangle \otimes \vert \phi^{\downarrow}_{\beta}
\rangle. 
\end{equation}
Any lowest energy state of $H'$ with $S^z_{tot}= {1\over 2}(N_t -2N_0)$ can be
expanded as
\begin{equation}
\vert \Psi(W)\rangle  = \sum_{\alpha,\beta} W_{\alpha\beta}{\bf T} \vert
\phi^{\uparrow}_{\alpha} \rangle \otimes \vert \phi^{\downarrow}_{\beta}
\rangle .
\end{equation}
The coefficients $\{ W_{\alpha\beta}\}$ can be regarded as a square matrix and
chosen as hermitian  since the transformed Hamiltonian ${\bf T} H' {\bf
T}^{\dagger}$ possesses spin up-down symmetry.

The proof of the positive definiteness of W is straightforward, but a little tedious
according to the reflection positivity approach in the spin space \cite{Lieb89,Shen96a}. The
variational energy of $H'$ is
\begin{eqnarray}
E(W) &\equiv & \langle\Psi(W)\vert H'\vert\Psi(W)\rangle \nonumber \\
    &=& \sum_{\sigma} 2 Tr(W^2 P)-\sum_{i\in\wedge} U_i Tr( W V_i W V_i) \nonumber \\
    &-& \sum_{i\in\wedge_d, \alpha_i} 
        \vert J_{i}\vert Tr(WV_{cdi\alpha_i}WV_{cdi\alpha_i}^{\dagger})\nonumber \\ 
    &-& \sum_{ij\in \wedge_d,\alpha_i,\alpha_j} \vert K_{ij}\vert
Tr(WV_{di\alpha_i j\alpha_j}WV_{di\alpha_ij\alpha_j}^{\dagger})\nonumber \\
    &-& \sum_{i\in \wedge_d,\alpha_i,\alpha'_i}\lambda_i Tr( W V_{di\alpha\alpha'} W V_{di\alpha\alpha'})\nonumber \\
    &+& \sum_{i\in \wedge_d}\lambda_i s_i(s_i +1),
\end{eqnarray}
where
\begin{eqnarray}
(P)_{\alpha\beta} &=&\langle \phi_{\alpha}^{\sigma}\vert
\sum_{ij\in \wedge}t_{ij}c^{\dagger}_{i,\sigma}c_{j,\sigma}\nonumber \\
&+& \sum_{i\in\wedge_d,\alpha_i} {J_i\over 2} (n_{i,\sigma} -{1\over 2})
(n_{i\alpha_i,\sigma} - {1 \over 2}) \nonumber \\
&+& \sum_{ij\in\wedge_d,\alpha_i,\alpha_j} 
{K_{ij}\over 2} (n_{i\alpha_i,\sigma} -{1\over 2})
(n_{j\alpha_j,\sigma} - {1 \over 2}) \nonumber \\
&+& \sum_{i\in\wedge_d,\alpha_i,\alpha'_i} {\lambda_i\over 2} 
(n_{i\alpha_i,\sigma} -{1\over 2})
( n_{i\alpha'_i,\sigma} - {1 \over 2}) \vert\phi^{\sigma}_{\beta}\rangle,
\end{eqnarray}
and
\begin{eqnarray}
( V_i )_{\alpha\beta} &=& \langle \phi_{\alpha}^{\sigma}\vert(n_{i\sigma} -
{1\over 2})\vert\phi^{\sigma}_{\beta}\rangle, \\
( V_{cdi\alpha_i} )_{\alpha\beta} &=& \langle \phi_{\alpha}^{\sigma}\vert
c_{i\sigma}d^{\dagger}_{i\sigma}\vert\phi^{\sigma}_{\beta}\rangle, \\
(V_{di\alpha_ij\alpha_j})_{\alpha\beta} &=&\langle \phi_{\alpha}^{\sigma}\vert
d_{i\alpha_i,\sigma}d^{\dagger}_{j\alpha_j\sigma}\vert\phi^{\sigma}_{\beta}\rangle,\\
(V_{di\alpha_i\alpha'_i})_{\alpha\beta} &=&\langle \phi_{\alpha}^{\sigma}\vert
d_{i\alpha_i,\sigma}d^{\dagger}_{i\alpha'_i,\sigma}\vert\phi^{\sigma}_{\beta}\rangle.
\end{eqnarray}
As $W$ is hermitian and can be decomposed as $W= V^{\dagger} DV$ where 
$V^{\dagger}V = 1$  and $D$ is a diagonal matrix with the diagonal elements
$d_{\alpha}$, denoted by $D=
diag\{ d_{\alpha}\}$. Denote $\vert W \vert =  V^{\dagger} \vert D\vert V$
where $\vert D\vert = diag\{ \vert d_{\alpha}\vert\}$. $\vert W\vert$ is at
least positive semidefinite as all its eigenvalues are non-negative. When all
$d_{\alpha} \neq 0$, $\vert W\vert$ is positive definite. As
\begin{eqnarray}
Tr(W^2 P) = Tr(\vert W\vert^2 P);\\
 Tr( W X W X^{\dagger}) \leq Tr( \vert W\vert X \vert W\vert X^{\dagger})
\end{eqnarray}
and  all $U_i\geq 0$ we have
\begin{equation}
E(W) \geq E(\vert W\vert).
\end{equation}
If $\vert \Psi(W)\rangle$ is the lowest energy state, $\vert \Psi(\vert
W\vert)\rangle$ must be also the lowest energy state in the variational principle. This indicates that 
one of the 
lowest energy states of $H'$ if degenerate is at least positive semidefinite. 
Suppose $\vert W\vert $ is positive semi-definite (not positive definite)
and there must exist one
non-zero vector $V_0$ such that $\vert W\vert V_0 =0$. From the Schr\"odinger equation
for $\vert W\vert$ and using the variational principle, we have 
\begin{eqnarray}
\vert W\vert N_{ci} V_0&=&0,~ ~   \makebox{if $U_i>0$}, \label{d}\\
\vert W\vert H_0 V_0&=&0,\\
\vert W\vert V_{cdi\alpha_i} V_0&=&0,~ ~   \makebox{if $J_i\neq 0$}, \\
\vert W\vert V_{di\alpha_i j\alpha_j} V_0&=&0,~ ~   \makebox{if $K_{ij}\neq 0$},\\
\vert W\vert V_{di\alpha_i\alpha_i} V_0&=&0,  \label{e}
\end{eqnarray}
where 
\begin{eqnarray}
(N_{ci})_{\alpha\beta} &=& \langle\phi^{\sigma}_{\alpha}\vert
n_{i\sigma}\vert\phi^{\sigma}_{\beta}\rangle,\\
(H_0)_{\alpha\beta} &=& \sum_{\langle ij\rangle\in \wedge}t_{ij}\langle\phi^{\sigma}_{\alpha}\vert
c^{\dagger}_{i\sigma}c_{j\sigma}\vert\phi^{\sigma}_{\beta}\rangle.
\end{eqnarray}
Eq. (36) is obtained from the single occupancy condition of $d$-fermions.
When all $U_i>0$ and $J_i\neq 0$, the generalised lattice is connected through
$t_{ij}$ and $J_i$. In this case, we can show that for any non-zero 
vector $V_{any}$ we have $\vert W\vert V_{any} =0$ by successively using
Eqs. (\ref{d}-\ref{e}). As $\vert W\vert$ is an non-zero matrix, it is
impossible that all of its eigenvalues are equal to zero. This indicates that
$\vert W\vert$ is positive definite and non of its eigenvalues are equal to zero. If there are two positive definite states,
for instance $W_1$ and $W_2$, constructing the state $W_1 + c W_2$ ($c$ is a
non-zero constant) and repeating the same procedure above we can show
that $W_1 = W_2$ or $W_1=-W_2$. Thus we conclude that the lowest energy state
in each subspace $S^z_{tot}$ is non-degenerate and positive 
definite. This conclusion holds even if some of $J_i =0$ but the generalised
lattice is connected through $t_{ij}$, $J_i$ and $K_{ij}$. 
However we cannot show the non-degeneracy of the lowest energy state in each
subspace if the generalised lattice is not
connected or  not all $U_i>0$ except for some special cases. 
A detailed proof for the case of $s=1/2$ is seen in Ref.\
\onlinecite{Shen96a}. In principle we have reduced the Kondo model with
large spin $\alpha_i$ to the model with spin 1/2. The proof for the case of
$1/2$ can be applied to the case of large spin. 

Since the lowest energy state is positive definite on the basis we choose, 
a theorem on a positive definite or semidefinite state \cite{Shen96a,Shen96b} can be applied to show
Theorem (ii-iv).  The theorem states:

Given  a positive semidefinite state $\vert \Phi\rangle$ on a bipartite lattice
with $N_A$ and $N_B$, then\\
1). If the state is
an eigenstate of the total spin, the eigenvalue $S$ is $\vert N_A-N_B\vert/2$ if $\vert
S^z\vert \leq \vert N_A - N_B\vert /2$, and $\vert S^z\vert$ otherwise;\\
2). The transverse spin-spin correlation function (for fermions with spin
$1/2$) obeys 
\begin{equation}
\langle \Phi\vert {\bf S}_i^+ \cdot {\bf S}_j^-\vert \Phi\rangle =
\epsilon(i)\epsilon(j)C_{ij} 
\end{equation}
where $C_{ij} \geq 0$ ( the equality only holds possibly in a positive
semidefinite, not definite state).

In our case, the difference of the total numbers of
the two generalised sublattice $\cal{A}$ and $\cal{B}$ sites is 
$\sum_{i\in\wedge} \epsilon(i) -
\sum_{i\in \wedge_d} {J_i \over \vert J_i\vert} 2s_i \epsilon(i)$. We get
Theorem (ii) and (iv) combining the theorem above
and the positive definiteness of
the lowest energy state. Theorem (iii) is also obtained as we have decomposed
the spin ${\bf S}_i$ into $2s_i$ $1/2$-spins and the whole system is still on a
generalised bipartite lattice.
For example,
\begin{eqnarray}
\langle\Psi\vert {\bf S}_i^+ \cdot {\bf S}_j^-\vert\Psi\rangle
&=& \sum_{\alpha_i,\alpha_j} \langle\Psi\vert 
d^{\dagger}_{i\alpha_i, \uparrow}d_{i\alpha_i, \downarrow}
d^{\dagger}_{i\alpha_i, \downarrow}d_{i\alpha_i, \uparrow}\vert\Psi\rangle
\nonumber \\
&=&\epsilon(i)\epsilon(j)F_{ij}
\end{eqnarray}
where
\begin{equation}
F_{ij} = \sum_{\alpha_i, \alpha_j} Tr(W^{\dagger}V_{di\alpha_ij\alpha_j}W
(V_{di\alpha_ij\alpha_j})^{\dagger}).
\end{equation}
$F_{ij} > 0$ when the state is positive definite, and $\geq 0$ when the state
is positive semidefinite.

In summary, we have provided several theorems on the ground state
properties of the Kondo model at half-filling and 
in the case with large spin. 
The uniqueness and the total spin in the ground state are found. Furthermore
we have also investigated the spin-spin correlation in the system. The
co-existence of both antiferromagnetic  and ferromagnetic long-range
correlations is proved, which is similar to the case of spin 1/2. 
Hopefully, these exact results are
useful for understanding this highly correlated electron-impurity lattice.

I would like to thank Prof. P. Fulde
for conversations.  I also wish to thank Dr. D. F. Wang for fruitful
discussions and  for his encouragement. 
This work was supported by the Alexander von Humboldt foundation of Germany.

\end{multicols}
\end{document}